# Sequential Optimal Placement of Distributed Photovoltaics using Downstream Power Index

Mir Hadi Athari, Charles Yang, Zhifang Wang
Department of Electrical and Computer Engineering
Virginia Commonwealth University
Richmond, Virginia, USA
Email:{atharih},{zfwang}@vcu.edu

*Abstract*—The optimization of the size and location of Photovoltaic (PV) Distributed Generation (DG) is a method for reducing distribution networks loss, cutting costs for utilities, and integrating renewable energy into the power grid. However, this optimization problem is a difficult mixed continuous discrete problem that is difficult to solve efficiently. Herein, we propose a novel Sequential PVDG Placement algorithm which utilizes a Downstream Power Index (DPI) for PV allocation. We compare Sequential PVDG Placement results with those of Shuffled Frog Leaping Algorithm (SFLA), a memetic heuristic algorithm. Our algorithm demonstrates similar accuracy and drastically less computation time compared to SFLA. Given its high accuracy and low computation time, sequential PVDG allocation algorithm may be useful for rapid online power dispatch, long-term planning, and microgrid operations under high penetration of renewable sources.

*Index Terms*-- Distributed Generation, Photovoltaics, Shuffled Frog Leaping Algorithm, Optimal Placement

## I. INTRODUCTION

A major concern for energy utilities is the power loss resulting from vast transmission distances. Meanwhile, climate change driven by carbon-emissions is a pressing issue for the global community. Recently, over 180 nations gathered to sign the Paris Accords, an international agreement to develop cleaner energy sources. As a result, there has been a rapid development in Photovoltaic Distributed Generation (PVDG), which provides renewable electric energy that is generated at the load site.

However, placing PVDG units within a radial network without considering optimal sizing and location may lead to deleterious effects, such as overvoltage fluctuations [1], active and reactive power regulation [2],and reverse power flow [3]. These phenomena can decrease the lifetime of voltage regulators [4], lead to loss of coordination between protection devices [5], and prevent the detection of faults [6]. In addition, rapid changes solar insolation can lead to power fluctuations, load imbalances, and undervoltage [7]. Therefore, there is a great incentive to optimize installation parameters, i.e. size and location, of PVDG. A variety of methods have been proposed and tested throughout literature. Viral and Khatod used an analytical method to find the optimal size and location of DG, assuming load was constant for the system [8]. Other analytical approaches such as sensitivity analysis and linear or quadratic approximations have been used as well [9], [10]. Population-based algorithms such as Differential Evolution [11], [12], Particle Swarm Optimization [13], and Shuffled Frog Leaping Algorithm (SFLA) [14] are also popular algorithms for this optimization problem.

Based on the review of literature, there are currently two main approaches to tackling the optimization of the size and location of PVDG with respect to energy loss: analytical methods and algorithmic(artificial intelligence) [15]. Purely analytical methods often utilize computationally expensive functions like Jacobian matrices [16]. On the other hand, heuristic algorithms are efficient and accurate, but may become trapped in local optima, suffer in performance when the solution space is large, and often require multiple trials to find good solutions. Since real world power distribution networks often have large systems with hundreds or thousands of buses, the solution space explodes rapidly as we consider multi-bus PVDG installations, a result of the combinatorial nature of the optimization problem. Conventional analytical methods like sensitivity factors begin to become too computationally expensive while optimization algorithms fail to efficiently search the entire solution space. High computation cost also becomes unacceptable as the "smart grid" becomes more prevalent and operations such as online dispatch, microgrid, and islanding, become used more often. For example, in the near future, online dispatch of DERs to avoid reverse power flow in the microgrids while achieving minimum power loss requires algorithms with online operation capability and less complexity. Thus, there is an urgent need for optimization methods that can handle large solution spaces rapidly and accurately.

In this paper, we propose the Sequential PVDG Placement Algorithm, which takes a hybrid approach to minimizing the total power loss in the distribution network. We define the Downstream Power Index (DPI) for each bus solely based on the load profiles, line resistance, and an average solar insolation that takes advantage of the radial topology found in most distribution networks. Because DPI does not need to run power flow, we greatly reduce the solution space and the computation time simultaneously. In this way, more complex and realistic grid models are computationally within reach and network planners are more likely to be able to identify optimal solutions. To determine sizing, we utilize the total downstream flow of the optimal location to scale up the PVDG so that revere power flow would be avoided. For comparison, we ran a Modified Shuffled Frog Leaping Algorithm (MSFLA), an evolutionary algorithm noted for its good convergence [17], [18] and compared results.

The rest of the paper is organized as follows. Section II describes the multi-feeder load and photovoltaic model. Section III is problem formulation. Section IV presents the Sequential PVDG Placement Algorithm and MSFLA and Section V discusses results. Section VI concludes the paper with recommendations for future works.

## II. FEEDER AND PHOTOVOLTAIC PANEL MODELS

### A. Multifeeder Model

Two distribution networks are selected for simulations, IEEE-69 bus and a fully designed 119 bus system from [19]. The 69 bus system consists of a main feeder (F1) and 7 other branches (F2-F8), with 3 different types of load profiles: residential, commercial, and industrial. A diagram of this system is shown with the average load of each bus (kW) and a specified color code designating the type of load in Fig. 1. The network voltage level is 12.66 kV and the total average load is 5.3+3.8 MVA.

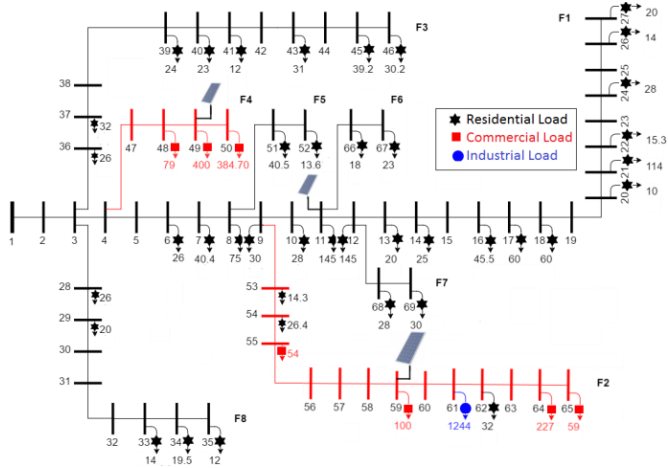

Figure 1. Schematic of redesigned 69 bus distribution network

The 119 bus distribution system is based on a local utility distribution system in Virginia, which feeds 1902 customers, with 1429 residential, 397 small commercial, and 76 large commercial/industrial classes. The network voltage level is 12.6 kV. The 119-bus system consists of a main feeder (A) and 6 other branches (B-F). The schematic diagram of the electrical network with color coded load types is shown in Fig. 2.

### B. PV module

The power generated (kW) by a PV array is given by [20] Solar Insolation data was gathered for Richmond, Virginia from the National Renewable Energy Lab (NREL) Physical Solar Model (PSM).

## III. PROBLEM FORMULATION

### A. Objective Function

The main objective of this model is to determine the optimal size and location of PVDG units within a radial multi-feeder distribution network that minimizes energy loss, subject to operational constraints. In a radial network, installing a PVDG at different buses along the feeder results into different level of loss reduction and requires different operational constraints. As shown in Fig. 3, the largest size of PVDG that can be installed without having reverse power flow to substation is at the beginning of the feeder while this configuration doesn't necessarily lead to maximum loss reduction due to smaller electrical distance ($Ed_1$) than further buses in the feeder. Conversely, the bus at the end of the feeder will have the largest electrical distance ($Ed_5$) but to avoid reverse power flow, the maximum PVDG size would be the smallest. This configuration won't necessarily result into the maximum loss reduction either. Therefore, the optimal point must be found somewhere in the middle of the feeder where PVDG size and electrical distance together give the highest reduction in network loss. We introduce a new index, called Downstream Power Index (DPI), to measure the potential of each bus of the network for loss reduction given maximum possible PVDG installed.

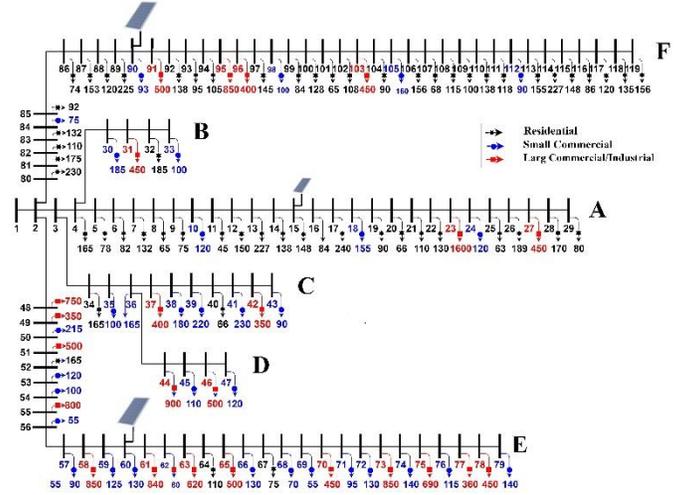

Figure 2. Schematic of redesigned 119 bus distribution network

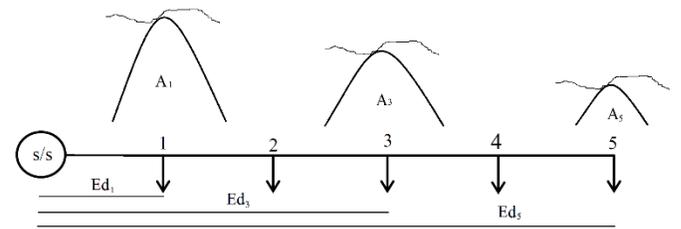

Figure 3. Concept of loss reduction for different configuration of PVDGs

The energy loss in a distribution network with total $L$ lines can be expressed as follows

$$E_{loss} = \sum_{t=1}^{24} \sum_{l=1}^{L} |i_L^t|^2 R_L \quad (1)$$

where $i_L^t$ is the current flowing through line $L$ at time $t$ and $R_L$ is the resistance of line $L$.

Thus, the objective function can be formulated as

$$\min_{\mathcal{L}_{PV}, P_{PV}^{max}} f = E_{loss} \quad (2)$$

Subject to:
$$f(P_L, P_{PV}, v|Y_{bus}) = 0 \quad (3)$$

$$i = h(v|Y_{bus}) \quad (4)$$
$$P_{PV}(t) = g(P_{PV}^{max}, I(t)) \quad (5)$$
$$i_{ij}^t \geq 0 \quad \forall\, i < j \quad (6)$$
$$0.95 \leq |v_i| \leq 1.05 \quad (7)$$
$$\sum \mathcal{L}_{PV} = N^{max} \quad (8)$$

where $\mathcal{L}_{PV} = [\ell_1, \ell_2, \ldots, \ell_n]^T$, $\ell_i \in \{0,1\}$ is the PVDG location vector, $P_{PV}^{max}$ is the PVDG maximum capacity vector, $Y_{bus}$ is the network admittance matrix, $i$ is vector of bus injected current, $v$ is bus voltage vector, $I(t)$ is solar insolation at time $t$, $i_{ij}^t$ is current flowing from bus $i$ to $j$ at time $t$, and $N^{max}$ is the maximum number of PVDG installations. Note that Eqs. (3) and (4) are network constraints enforced by AC power flow and network operation constraints, respectively. Eq. (5) is the PVDG installation constraint.

## IV. OPTIMAL PLACEMENT OF PVDG

### A. Sequential PVDG Placement Algorithm

The Sequential PVDG Placement Algorithm uses the Downstream Power Index (DPI) to identify the optimal location for PVDG. DPI is an analytical method that only requires the load profiles, resistance of lines, network incidence matrix, and the average solar insolation profile to identify optimal locations, thus greatly reducing the solution space without needing to run the computationally expensive power flow equations or utilize complex linear algebra methods. First, our algorithm reads the network topology matrix to automatically determine which buses are at the downstream of each bus. Next, we find the total power downstream of each bus with eq. (9).

$$P_{downstream}(i,t) = P_{Load}(i,t) + \sum_{j=1}^{k} P_{Load}(j,t) \quad (9)$$

where $P_{downstream}(i,t)$ is the downstream power connected to bus $i$ at time $t$ and $P_{Load}(j,t)$ is the load power at bus $j$ that is connected to buses at the downstream of bus $i$.

The contribution of PVDG units in network loss reduction during a 24 hour period depends on their output power availability. In other words, PVDGs help minimize the network loss from sunrise to sunset which we call it Feasible Optimization Interval (FOI). According to eq. (6), to achieve the highest loss reduction without having reverse power flow, we have to scale the capacity of the PVDG according to downstream power of each bus. Hence, we use a standard yearly average solar insolation profile in the calculation of DPI for the study area (Fig. 4). Using the PV panel model proposed in [19], the PV panel output (kW) is calculated for the average solar insolation ($P_{PV}^{panel}$). Note that, here we assume that the networks under study cover a limited geographic area and the uniform solar insolation assumption is rational. Next, we use the downstream power at the panel peak output power to scale the capacity of hypothetical PVDG at each bus:

$$P_{PV}(i,t) = \left(\frac{P_{downstream}(i, t_{max}^{panel})}{\max(P_{PV}^{panel})}\right) \times P_{PV}^{panel}(t) \quad (10)$$

where $t_{max}^{panel}$ is the time that the PV panel generates its nominal output.

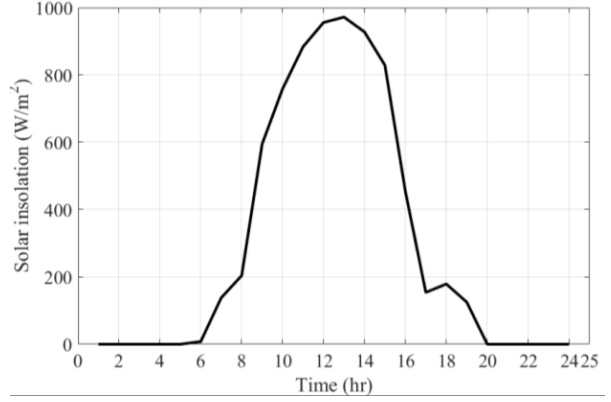

Figure 4. Yearly average solar insolation for study areas

Next, we approximate the downstream PV current using Ohm's Law that will contribute in network loss reduction:

$$I_{downstream}^{PV}(i,t) \approx \frac{P_{PV}(i,t)}{V_{base}} \quad (11)$$

where $V_{base}$ is the base voltage level of the distribution network (12.6 kV).

We then define the electrical distance of some bus $i$ from the substation as the sum of the resistances of the $m$ lines that connect bus $i$ to the substation.

$$Ed(i) = \sum_{l=1}^{m} R_s(i,l) \quad (12)$$

where $R_s$ is the resistance of lines connecting bus $i$ to the substation.

Finally using the electrical distance and downstream current, we are able to define the DPI as an indicator of the potential of each bus to minimize the total network loss.

$$DPI(i) = \sum_{t=1}^{24} I_{downstream}^2(i,t) \times Ed(i) \quad (13)$$

We then normalize the DPI for each bus relative to the largest DPI value: $\overline{DPI}(i) = \frac{DPI(i)}{DPI^{max}}$.

The rationale for this analysis is as follows. Based on our problem formulation, the optimal location is the bus that has the largest amount of power downstream that must travel over the longest electrical distance, as the power will no longer have to travel over those lines but will instead be provided by PV locally. Because the transmission losses from any given bus to the end of the feeder are negligible compared to the load delivered, we are able to use the power downstream as an accurate approximation of grid dynamics. In addition, by considering the downstream power in DPI, we also take into account our constraints: reverse power flow will only occur when the PV power delivered is greater than the power downstream and overvoltage only occurs after reverse power flow. Finally, we determine how well the max PV scales to the load profile, which is essentially a measure of how well the profiles fit with each other.

After finding the optimal location, the size of the PVDG is assigned after determining the total downstream flow of optimal bus using power flow analysis. The size is selected so that the peak PV output equals the total power flowing downstream of the bus plus already existing load on the bus.

Finally, for *n* PV installations, Sequential PVDG Placement takes the optimal solution for *n-1* PV bus installations and reruns DPI and associated installed PVDG size. For 2 bus installation, Sequential PVDG Placement takes the optimal solution for 1 PV installation, reruns power flow with the PV installed and identifies the best size and location for the 2nd bus. Our implicit assumption that the optimal solution for *n* PV installations is dependent on the optimal solutions of the 1…*n-1* PV installations is based on our group's previous observations. Further investigation into this phenomena is certainly worthy of merit and one we will be considering in the future. However, based on our results in Section V, this methodology is defensible in the excellent results and greater computational efficiency it returns. The flow chart of the proposed sequential PVDG placement algorithm is shown in Fig. 5.

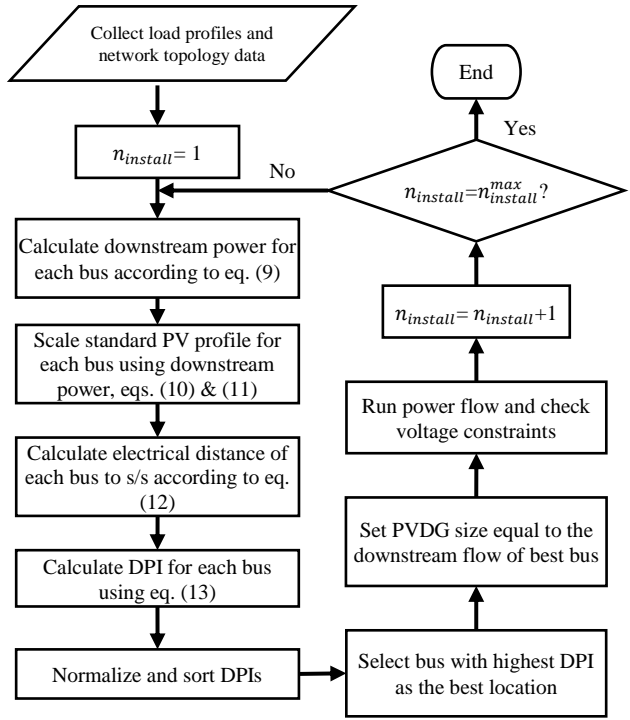

Figure 5. Flow chart of Sequential PVDG placement

### B. Modified Shuffled Frog Leaping Algorithm

SFLA combines ideas from shuffled complex evolution for global search and particle swarm optimization for local search [18]. We describe SFLA and our modifications here.

First, a random population of *P* vectors, or frogs, are initialized. These vectors are then ranked by their fitness and then sorted into *m* number of memeplexes, where the best frog, $X_g$, goes into the 1st memeplex, the second best frog into the 2nd memeplex, until the $m^{th}$ ranked frog goes into the $m^{th}$ memeplex, and then the *m*+1 rank frog goes to the first memeplex, and so on. The worst and best frog within each memeplex, $X_w$ and $X_b$ respectively, are then identified and $X_w$ is evolved into $X_{wnew}$ according to the following heuristic.

$$X_{wnew} = X_w + D \quad (14)$$
$$D = r \times C_1 \times (X_b - X_w) + r \times C_2 \times (X_g - X_w) \quad (15)$$

where $D$ is the leaping step, $r$ is a random number such that $0 \le r \le 1$, and $C_1, C_2 \in [1,2]$. The scaling constants $C_1, C_2$ help $X_w$ to "accelerate" past $X_b$ and $X_g$ to effectively search around the optimum, rather than linearly approach them. Each $X_w$ frog in each memeplex is modified according to the leaping step 90% of the time. In order to introduce diversity, we discard $X_w$ and create a new frog based on mutation and diversity for 10% of the time. We choose two random frogs, $X_1$ and $X_2$, such that $X_1$ is in the top half ranking in terms of fitness and $X_2$ is from the top third ranking in terms of fitness. This elitist mutation of $X_1$, $X_2$, and $X_g$ introduces diversity into each memeplex.

$$X_{wnew} = r \times C_3 \times (X_g - X_1) + X_2 \quad (16)$$

Thus the worst frog moves toward the local optimal solution and explores the solution space around it or it is mutated. After this, the frogs are ranked by their fitness again and sorted into new memeplexes, thus promoting diversity. At the end of each iteration, $X_g$ is identified. After a set number of iterations, the algorithm ends. SFLA parameters used in this paper are described in Table I.

TABLE I. SFLA PARAMETERS

| Shuffled Frog Leaping Algorithm Parameters | | | |
|---|---|---|---|
| # of memeplexes | 10 | Max # of iterations | 200 |
| # of frogs per memeplex | 10 | Max Size of PV (MW) | 5.20 |
| Total number of frogs | 100 | # of bus locations | 69, 119 |

## V. RESULTS

The Sequential PVDG Placement Algorithm and MSFLA were both run on the 69 bus and 119 bus systems. MSFLA was run for 5 trials and the average cost and runtime were determined. It should be noted that heuristic algorithms often require multiple trials or runs to identify global optima, while Sequential PVDG Placement Algorithm needs only to be run once. Table II and III show the bus rankings according to DPI in the Sequential PVDG Placement algorithm for 69 and 119 bus systems, respectively. Only top 5 buses are shown for comparison with the optimal solutions found by MSFLA.

TABLE II. BUS RANKINGS ACCORDING TO PDI FOR 69 BUS SYSTEM

| First PVDG | | Second PVDG | | Third PVDG | |
|---|---|---|---|---|---|
| Bus # | $\overline{PDI}$ | Bus # | $\overline{DPI}$ | Bus # | $\overline{DPI}$ |
| 59 | 100 | 11 | 100 | 49 | 100 |
| 58 | 97.58254317 | 10 | 94.89833025 | 51 | 73.63009138 |
| 57 | 91.35459867 | 8 | 86.74969753 | 7 | 64.79671441 |
| 61 | 90.19675815 | 12 | 85.82435056 | 8 | 60.01263126 |
| 60 | 86.80031432 | 7 | 82.5764199 | 6 | 38.05523171 |

TABLE III. BUS RANKINGS ACCORDING TO PDI FOR 119 BUS SYSTEM

| First PVDG | | Second PVDG | | Third PVDG | |
|---|---|---|---|---|---|
| Bus # | $\overline{PDI}$ | Bus # | $\overline{DPI}$ | Bus # | $\overline{DPI}$ |
| 60 | 100 | 90 | 100 | 15 | 100 |
| 59 | 97.98569016 | 91 | 99.87106349 | 17 | 98.45080645 |
| 61 | 97.58641344 | 15 | 91.92992938 | 18 | 97.43036847 |
| 90 | 76.18212023 | 17 | 90.50575685 | 16 | 97.35864668 |
| 91 | 76.08389366 | 18 | 89.56766893 | 21 | 96.30918777 |

Tables IV and V show the optimal placement results based on our proposed sequential algorithm and MSFLA, respectively. The optimal locations are shown on Figs. 1 and 2.

It is found that our proposed sequential algorithm has comparable performance in terms of total energy loss reduction with the best solution of MSFLA for both 69 and 119 bus systems and the solution from our algorithm is within 1% of the optimal solution. However, the computation time for the proposed method is 755 times and 1048 times faster than MSFLA for 69 and 119 bus systems, respectively. This is a result of the DPI, which can identify an optimal location very rapidly, eliminating the need to search a large solution space. Also, note that the average total energy loss achieved by MSFLA for 5 runs, is larger than that of our algorithm. This shows that heuristic algorithms tend to get trapped in local optima, hence requiring several runs to actually find the global optima which basically makes them even more computationally expensive.

TABLE IV.    RESULTS FOR SEQUENTIAL PVDG PLACEMENT ALGORITHM

| Sequential PVDG Placement Algorithm | | | |
|---|---|---|---|
| 69 Bus System | | | |
| Location (bus) | 59 | 11 | 49 |
| Size (MW) | 1.95 | 0.80 | 0.74 |
| Runtime(sec) | 1.1704 | | |
| Total energy loss (kWh) | 3672.6 | | |
| 119 Bus System | | | |
| Location (bus) | 60 | 90 | 15 |
| Size (MW) | 4.73 | 4.13 | 2.96 |
| Runtime(sec) | 2.2045 | | |
| Total energy loss (kWh) | 21000 | | |

TABLE V.    RESULTS FOR MSFLA

| Modified Shuffled Frog Leaping Algorithm | | | |
|---|---|---|---|
| 69 Bus System | | | |
| Best Location (bus) | 61 | 11 | 49 |
| Best Size (MW) | 1.75 | 0.80 | 0.74 |
| Average Runtime (sec) | 884 | | |
| Best cost (kWh) | 3664.7 | | |
| Average Total Energy Loss (kWh) | 3720 | | |
| 119 Bus System | | | |
| Best Location (bus) | 61 | 91 | 15 |
| Best Size (MW) | 4.60 | 4.02 | 2.96 |
| Average Runtime (sec) | 2310 | | |
| Best cost (kWh) | 20968.8 | | |
| Average Total Energy Loss (kWh) | 21355 | | |

When we increase the size of the network to 119 buses, for fixed algorithm parameters, the run time of MSFLA nearly triples, as the solution space becomes much larger, but the runtime of Sequential PVDG Placement Algorithm is 2.2 seconds. This shows that runtime for heuristic algorithms scales almost exponentially with respect to network size while the runtime of the Sequential algorithm increases linearly as system size scales up.

The reason that MSFLA performs slightly better in terms of total loss reduction is the difference in the optimal location for the first and second PVDGs. For example in 69 bus system, our algorithm finds bus 59 as the best location, while the optimal location is at bus 61, two buses downstream of 59 and at the same branch. This is due to approximation in the calculation of PDI, since we neglect the line loss in the downstream power to avoid power flow calculation. Note that, for both systems, the optimal locations are always in the top 5 buses according to PDI, thus by running power flow calculation we may find the optimal location and improve the performance of the sequential algorithm. However, the computation time of our algorithm is thus far unparalleled, based on our literature review. Our blisteringly fast algorithm can thus also be applied to problems outside of the traditional domain of power grid planning, such as online power dispatch and microgrid operations under high penetration of renewable energy sources, both of which require much faster computational speeds than can be currently delivered by conventional techniques.

Another point worthy of notice is that for both systems, the optimal bus is located somewhere in the middle of the branches of the main feeder which is a result of radial topology and validates the rationale behind our analytical method.

Figs. 6 and 7 show the voltage profiles of the branches before and after installation of PVDGs for 69 and 119 bus systems, respectively. For both systems, installing PVDG has boosted up the voltage profile along the branch during the hours of the day with available solar power output. Therefore, we have flatter voltage profile and near nominal value near the end of the feeder while we see a considerable voltage drop for the original system that can cause damage to voltage sensitive devices. In order to have better voltage profile for all hours of the day, the use of storage devices together with PVDGs is recommended so that extra energy generated by PVDG can be stored for later use during nightfall.

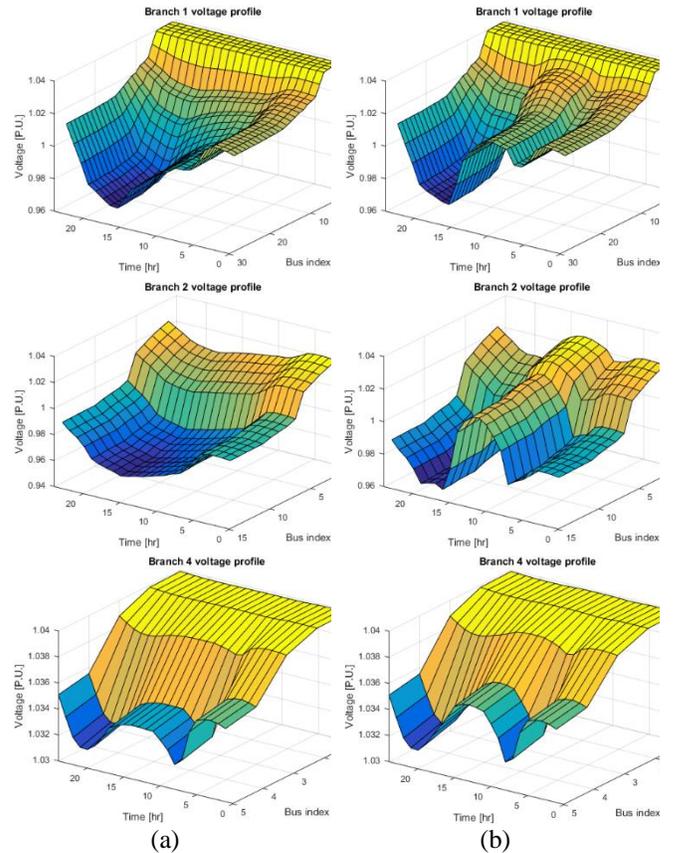

Figure 6.  Voltage profiles for 69 bus system: a) before PVDG installation, b) after PVDG installation

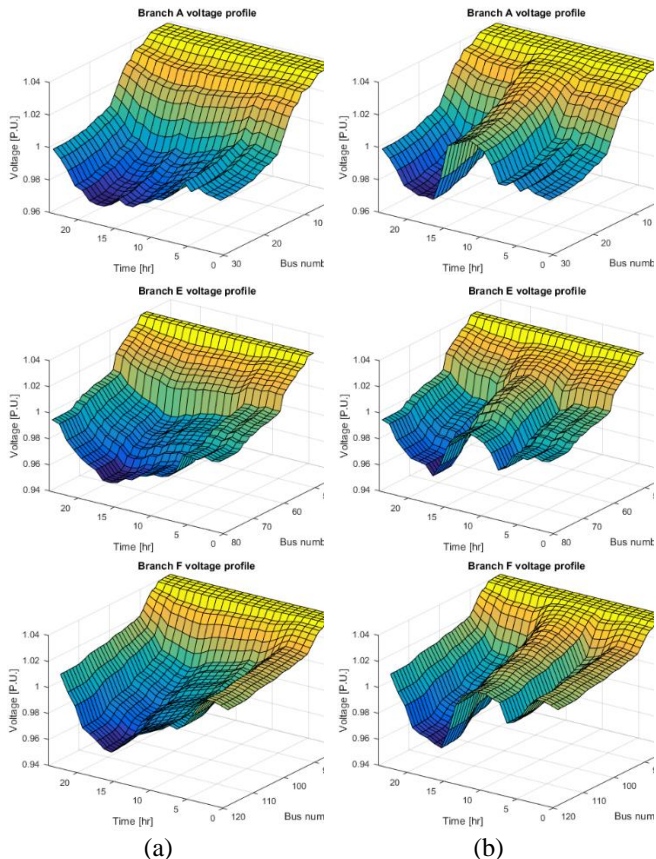

Figure 7. Voltage profiles for 119 bus system: a) before PVDG installation, b) after PVDG installation

## VI. Conclusion and Future Works

Optimal placement of PVDG units with the objective of minimizing the total energy loss of the distribution network based on an analytical method is presented in this paper. We propose the Sequential PVDG Placement Algorithm that is based on the observation that the optimal solution for multi bus installation is dependent on previous optimal solutions. Our Downstream Power Index (DPI) is a rapid analytical method that can identify optimal locations based solely on network topology, load profiles, average solar insolation, and line resistance. Our results demonstrate that Sequential PVDG Placement Algorithm delivers order of magnitude enhancement in computation time and comparable results in accuracy. We believe that this method has potential applications in microgrid operations and online power dispatch under high penetration of renewables, which are currently too time-demanding for conventional techniques in literature. Finally, the underlying rationale in DPI can be used to improve other algorithms and may provide possible alternatives to power flow in many electrical engineering problems. Further work includes application of the proposed DPI method to dispatch multiple PVDGs in a microgrid to avoid reverse power flow while achieving the minimum power loss in the grid.

## References

[1] H. Alatrash, R. A. Amarin, and C. Lam, "Enabling Large-Scale PV Integration into the Grid," in *2012 IEEE Green Technologies Conference*, 2012, pp. 1–6.

[2] M. E. Raoufat, A. Khayatian, and A. Mojallal, "Performance Recovery of Voltage Source Converters With Application to Grid-Connected Fuel Cell DGs," *IEEE Trans. Smart Grid*, pp. 1–8, 2016.

[3] J. R. Aguero and S. J. Steffel, "Integration challenges of photovoltaic distributed generation on power distribution systems," in *2011 IEEE Power and Energy Society General Meeting*, 2011, pp. 1–6.

[4] S. Teleke, F. Jahanbakhsh, F. Katiraei, and J. Romero Aguero, "Analysis of interconnection of photovoltaic distributed generation," in *2011 IEEE Industry Applications Society Annual Meeting*, 2011, pp. 1–6.

[5] M. E. Baran, H. Hooshyar, Z. Shen, and A. Huang, "Accommodating High PV Penetration on Distribution Feeders," *IEEE Trans. Smart Grid*, vol. 3, no. 2, pp. 1039–1046, Jun. 2012.

[6] H. Cheung, A. Hamlyn, L. Wang, C. Yang, and R. Cheung, "Investigations of impacts of distributed generations on feeder protections," in *2009 IEEE Power & Energy Society General Meeting*, 2009, pp. 1–7.

[7] M. Farhoodnea, A. Mohamed, H. Shareef, and H. Zayandehroodi, "Power quality impact of grid-connected photovoltaic generation system in distribution networks," in *2012 IEEE Student Conference on Research and Development (SCOReD)*, 2012, pp. 1–6.

[8] R. Viral and D. K. Khatod, "An analytical approach for sizing and siting of DGs in balanced radial distribution networks for loss minimization," *Int. J. Electr. Power Energy Syst.*, vol. 67, pp. 191–201, 2015.

[9] A. Al Ameri, C. Nichita, H. Abbood, and A. Al Atabi, "Fast Estimation Method for Selection of Optimal Distributed Generation Size Using Kalman Filter and Graph Theory," in *2015 17th UKSim-AMSS International Conference on Modelling and Simulation (UKSim)*, 2015, pp. 420–425.

[10] S. Essallah, A. Bouallegue, and A. Khedher, "Optimal placement of PV-distributed generation units in radial distribution system based on sensitivity approaches," in *2015 16th International Conference on Sciences and Techniques of Automatic Control and Computer Engineering (STA)*, 2015, pp. 513–520.

[11] M. Md Rasid, M. Junichi, and T. Hirotaka, "Simultaneous determination of optimal sizes and locations of Distributed Generation units by Differential Evolution," in *2015 18th International Conference on Intelligent System Application to Power Systems (ISAP)*, 2015, pp. 1–6.

[12] B. Mahdad and K. Srairi, "Adaptive differential search algorithm for optimal location of distributed generation in the presence of SVC for power loss reduction in distribution system," *Eng. Sci. Technol. an Int. J.*, vol. 19, no. 3, pp. 1266–1282, 2016.

[13] N. Ameir Ahmad, I. Musirin, and S. I. Sulaiman, "Loss minimization of distribution system with photovoltaic injection using Swarm evolutionary programming," in *2013 IEEE 7th International Power Engineering and Optimization Conference (PEOCO)*, 2013, pp. 752–757.

[14] H. Doagou-Mojarrad, G. B. Gharehpetian, H. Rastegar, and J. Olamaei, "Optimal placement and sizing of DG (distributed generation) units in distribution networks by novel hybrid evolutionary algorithm," *Energy*, vol. 54, pp. 129–138, 2013.

[15] P. Prakash and D. K. Khatod, "Review of methods for optimal sizing and siting of Distributed Generation," in *2015 Annual IEEE India Conference (INDICON)*, 2015, pp. 1–6.

[16] Cuo Zhang, Y. Xu, ZhaoYang Dong, and Jin Ma, "A composite sensitivity factor based method for networked distributed generation planning," in *2016 Power Systems Computation Conference (PSCC)*, 2016, pp. 1–7.

[17] M. Eusuff, K. Lansey, and F. Pasha, "Shuffled frog-leaping algorithm: a memetic meta-heuristic for discrete optimization," *Eng. Optim.*, vol. 38, no. 2, pp. 129–154, Mar. 2006.

[18] A. Sarkheyli, A. M. Zain, and S. Sharif, "The role of basic, modified and hybrid shuffled frog leaping algorithm on optimization problems: a review," *Soft Comput.*, vol. 19, no. 7, pp. 2011–2038, Jul. 2015.

[19] H. Sadeghian, M. H. Athari, and Z. Wang, "Optimized Solar Photovoltaic Generation in a Real Local Distribution Network," in *2017 IEEE Power & Energy Society Innovative Smart Grid Technologies Conference (ISGT) (2017)*, 2017.

[20] M. H. Athari and M. M. Ardehali, "Operational performance of energy storage as function of electricity prices for on-grid hybrid renewable energy system by optimized fuzzy logic controller," *Renew. Energy*, 2016.